\documentclass[article,twocolumn,showpacs,superscriptaddress,amsmath,amssymb]{revtex4}

\usepackage{cancel}
\usepackage{maybemath}
\usepackage{xcolor}
\usepackage{graphicx}

\begin{document}

published as Astropart. Phys., 85, 43-49 (2016)\\

\title{3 neutrino mass experiments fit a strange 3 + 3 model,\\but will KATRIN reveal the model's unique 3-part signature?}

\author{R. Ehrlich}

\affiliation{George Mason University,\\ Fairfax, Virginia 22030, USA}
\email{rehrlich@gmu.edu}

\date{\today}

\begin{abstract}
Evidence is presented in support of an unconventional $3+3$ model of the neutrino mass eigenstates with specific $m^2>0$ and $m^2<0$ masses.  The two large $m^2>0$ masses of the model were originally suggested based on a SN 1987A analysis, and they were further supported by several dark matter fits.   The new evidence for one of the $m^2>0$ mass values comes from an analysis of published data from the $\emph{three}$ most precise tritium $\beta-$decay experiments.  The KATRIN experiment by virtue of a unique 3-part signature should either confirm or reject the model in its entirety.
\end{abstract}
\pacs{14.60.Pq, 14.60.St}
\maketitle

\section{Introduction}
The 3+3 model developed from an analysis of SN 1987A data that provided evidence for two heavy neutrino mass states that were inferred from the observed neutrino energies and arrival times.~\citep{Co1988, Eh2012,Eh2013}  Despite the conservative upper limit of 12 eV~\citep{Ar1987} on the $\nu_e$ rest mass from these data, ref.~\citep{Eh2012} showed that if one assumes that the emission times of the observed SN 1987A neutrinos have a spread of $\sigma_t \approx \pm 0.5 sec$ then they were all consistent with having masses of either $4.0\pm 0.5 eV$ or $21.4\pm 1.2 eV.$  Further evidence for these two masses was provided based on excellent fits to the dark matter halo profiles in both the Milky Way (21.4 eV), and four clusters of galaxies (4.0 eV).~\citep{Ch2014}

\subsection{Basis of the $3+3$ model}
Given the large masses noted above the neutrino mass states would have to be $\nu_L,\nu_R$ doublets to be consistent with the small $\Delta m^2$ seen in neutrino oscillation experiments -- see sect.~\ref{Osc} for more detail.  Furthermore, given two heavy $m^2>0$ doublets, a third singlet or doublet would need to have $m^2<0,$ and be a tachyon, so as to yield flavor state masses $m_i$ given by $m^2_i = \Sigma |U_{i,j}|^2 m_{j}^2$ that are all very small so as to satisfy the cosmological upper limit on their sum: $m_{tot}=\Sigma m_i <0.3 eV$~\citep{Ol2015} --  see sect.~\ref{Cosmology}.  Thus, the 3+3 model postulated two $\nu_L,\nu_R$ doublets of mass $m^2>0$ with splittings: $\Delta m^2_{12}=\Delta m^2_{sol}$ (solar) and $\Delta m^2_{34}=\Delta m^2_{atm}$ (atmospheric).  For the $m^2<0$ state the natural choice was a doublet with a splitting $\Delta m^2_{56}\approx 1 eV^2$ inferred from several oscillation experiments~\citep{Ko2011, Ka2015}.  Once it was realized that $\Delta m_{12}^2/m_1^2$ and $\Delta m_{34}^2/m_3^2$ had identical values it was suggested in ref.~\citep{Eh2013} that the natural choice of the mass for the $m^2<0$ doublet would be $m_5^2 \approx m_6^2 \approx -0.2 keV^2,$ so that all three doublets would have identical fractional splittings: $\Delta m^2/m^2=5.0\times 10^{-6}.$  

The choice of $-0.2 keV^2$ may have a theoretical justification beyond that of a highly suggestive numerical coincidence, namely one based on CPT, which if conserved would require $\nu$ and $\bar{\nu}$ states to have the same mass.  In ref.~\citep{Eh2013} it was assumed that the three $\nu_R$ states were sterile neutrinos, but if instead they should be active $\bar{\nu}$ states, the small and constant $\Delta m^2/m^2$ of each of the $\nu_L,\nu_R$ doublets might signal a small violation of CPT in the neutrino sector, which may be experimentally testable.~\citep{Sm2015}.  Moreover, this possibility would then imply a small degree of Lorentz violation,~\citep{Gr2002} which is required by many tachyon theories.  The $3+3$ model is clearly speculative, and its most controversial element is the claim of $m^2<0$ neutrinos.

\subsection{Stable $m^2<0$ tachyonic neutrinos with $v>c$?} 
In its conventional usage the imaginary mass of a tachyon field creates an instability leading to a spontaneous decay as in the Higgs mechanism,~\citep{Ge1974} with no $v>c$ propagation and a tachyon-free resulting state after decay.  Here, however, we are considering the more controversial idea of tachyons as stable $m^2<0$ particles having $v>c.$  Ever since such hypothetical particles were first proposed in 1962 by Bilaniuk, Deshpande, and Sudarshan,~\citep{Bi1962} probably only a small fraction of physicists such as Feinberg~\citep{Fe1967} and Recami~\citep{Re1978} have taken the subject seriously.  In fact, many physicists believe that stable $m^2<0$ particles with $v>c$ cannot exist on theoretical grounds.  However, in 1985 theorists Chodos, Hauser and Kostelecky proposed that neutrinos could be tachyons and wrote a Dirac-like equation for them.~\citep{Ch1985}  Since that time various other theorists have suggested modifications to the standard model or quantum field theory that could accommodate $m^2<0$ neutrinos.  These include small departures from Lorentz Invariance~\citep{Al2011}, or unitarity,~\citep{Je2012a}, assuming pseudo-Hermitian operators~\citep{Je2012b}, or a preferred reference frame,~\citep{Re1997} or invoking a new symmetry principle,~\citep{Ch2016} or Majorana mass-mixing with a sterile negative-metric field.~\citep{Ch2001}  Given the preceding theoretical work, the possibility of neutrinos as tachyons really needs to be considered an empirical question.  Extraordinary claims such as that of $m^2<0$ neutrinos, one part of the $3+3$ model, require extraordinary evidence.  The evidence in the next section probably falls short of that level, but surely if the results from the KATRIN experiment were to show the model's unique 3-part signature that would be truly extraordinary.

\section{Direct $\nu$ mass experiments}
KATRIN is the most precise neutrino mass experiment now in operation.  It has the goal of either setting an improved upper limit (by a factor of ten) on the effective mass of the electron neutrino or discovering its value through a measurement of the tritium $\beta-$decay spectrum near its endpoint.~\citep{Dr2013}  A subsidiary purpose is to seek evidence for possible sterile neutrino states given the resulting distortion they might cause to the spectrum.~\citep{Me2015}  

\subsection{Fitting the spectrum using the Kurie function}
The square of the Kurie function in $\beta-$decay can be written as~\citep{Gi2007}

\begin{equation}
K^2(E)=(E_0-E)\sum |U_{ej}|^2\sqrt{R((E_0-E)^2-m_j^2)}
\label{A1}
\end{equation}

where $E$ is the electron's kinetic energy, $E_0$ is the endpoint energy, and $m_j$ are the masses of the mass states (including possible sterile states) making up the $\nu_e$ flavor state, through $|\nu_e>=\sum U_{ej}|\nu_j>$.  $R$ is the Ramp function defined by $R(x)=x$ for $x\geq 0,$ and $R(x)=0$ for $x<0.$ Its use in Eq. ~\ref{A1} ensures that when $(E_0-E)^2-m_j^2$ is negative it is replaced by zero, so that the expression under the square root is never negative which cannot occur for $m^2<0$ neutrinos.

Under certain assumptions $K^2(E)$ describes part of the $\beta-$decay spectrum, and a fit to that spectrum in principle might allow us to determine the value of some of the $m_j.$   In effect, the result of the sum over j in Eq. ~\ref{A1}, is a sum of separate spectra for the individual $m_j$ with weights $|U_{ej}|^2.$  If one or more of the $m_j$ is sufficiently heavy and the mixing angle with the light states is not too small one might detect it based on a ``kink" in the spectrum at a distance $m_j$ below the endpoint  The usual procedure is to assume $m_1\approx m_2 \approx m_3=0,$ and fit the spectrum assuming only one sterile neutrino having an unknown mass ($m_4$).  To date several groups have carried out such searches and established upper limits on $m_4$ or more precisely the allowed region of the $U^2_{e4}$ versus $m_4$ plane that is consistent with their data.~\citep{Kr2013, Be2013, Be2014}

\subsection{Fitting the data to the 3 + 3 model\label{IIIB}}
The KATRIN experiment is also capable of testing the $3+3$ model through a unique three-part signature.  Recall that the model includes two unconventionally heavy $\nu_L,\nu_R$ doublets with $m_1\approx m_2=4.0\pm 0.5 eV$ and $m_3\approx m_4=21.4 \pm 1.2 eV.$ and a third doublet with $m^2_5\approx m^2_6=-0.2 keV^2.$  In what follows we shall make the assumption of the above values for the six $m_j$, and further that the magnitude of the effective mass of the $\nu_e$ flavor state is much less than 1 eV, so as to satisfy the upper limit on $\Sigma m$ from cosmology.  Thus, in calculating the spectrum for the $3+3$ model we have the two requirements that $\sum |U_{ej}|^2=1$ and $m^2_{eff}=\sum |U_{ej}|^2 m_j^2=0,$ and so we have only one free parameter in Eq. ~\ref{A1}, i.e.,  $U^2_{12}\equiv |U^2_{e1}|+|U^2_{e2}|.$   Thus, by choosing a value for $U^2_{12}$ we can obtain the two others: $U^2_{34}\equiv |U^2_{e3}|+|U^2_{e4}|,$ and $U^2_{56}\equiv |U^2_{e5}|+|U^2_{e6}|,$ and then use Eq.~\ref{A1} to calculate the spectrum.  We defer until later the question of whether the functional form of Eq.~\ref{A1} might not apply to $m^2<0$ neutrinos.  \emph {It is very important to be aware that fits to the spectrum that place restrictions on a single mass parameter, i.e., either $m_4$ or $m_{eff},$ are of limited relevance when doing fits involving three large pre-defined mass values.}  Specifically, even though the two most precise experiments to date require as an upper limit $m_{eff}<2.0 eV$ for the $\nu_e$ effective mass, as we shall see excellent fits can be found to the data using the large $3+3$ model masses.

\begin{figure}[t!]
\centerline{\includegraphics[angle=270,width=1\linewidth]{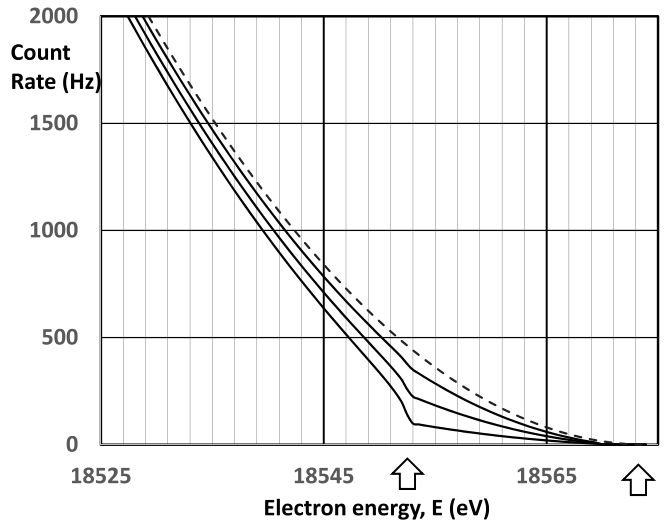}}
\caption{\label{fig3}
Calculated spectra using 3+3 model.  The plot shows the count rate versus neutrino kinetic energy $E$ using Eq. ~\ref{A1} within 50 eV of the endpoint (upward right arrow): (a) assuming all neutrinos to be massless (dashed curve), and (b) (solid curves) using values for the six neutrino mass states equal to the values defined in the text, and the mixing parameter $U^2_{12}=U^2_{e1}+U^2_{e2}$ taking the values 0.8 (top), 0.5 (middle), and 0.2(bottom) solid curves.  The left upward arrow shows the position of the kink in the three spectra occurring a distance $m_3=21.4 eV$ before the endpoint.  The vertical scale is arbitrary, and yields one count per second for $E_0-E =1eV$.}
\end{figure}

\subsection{Some 3+3 model spectra near endpoint}
We see, for example, in Fig.~\ref{fig3} the results under several assumed values for $U^2_{12}$ over the region for $E_0-E <50eV,$ with the endpoint assumed to lie at $E_0=18574 eV.$  The dashed curve is the expected spectrum for all $m_j=0,$ while the three solid curves correspond to $U^2_{12}=0.8,0.5$ and $0.2$ for the top, middle and bottom one.  On the scale of Fig.~\ref{fig3} there is only one noticeable feature in the three solid curves compared to the $m_j=0$ curve, i.e., the kink that occurs at $m_3=21.4 eV$ below the spectrum endpoint.  We see that, as expected, the amplitude of that kink is smaller the larger we make $U^2_{12},$ and hence the smaller we make $U^2_{34}.$

\begin{figure}[t!]
\centerline{\includegraphics[angle=270,width=1\linewidth]{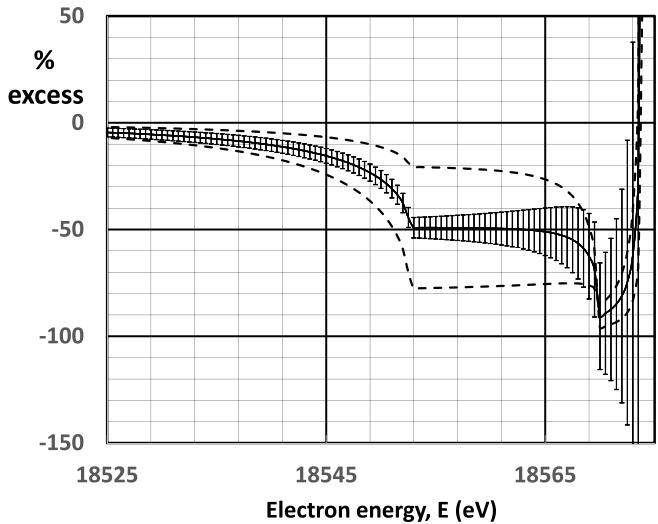}}
\caption{\label{fig4}
Percentage excess (or deficit) of 3 + 3 spectrum relative to the spectrum with all $m_j=0$ for each of the three cases of Fig.~\ref{fig3}.  The solid curve with error bars is for the middle case $U^2_{12}=0.5,$ and the upper (lower) dashed curves correspond to $U^2_{12}=0.8(0.2).$}
\end{figure}

There are in fact two other departures of the spectrum shape from the $m_j=0$ curve (not visible in Fig.~\ref{fig3}) that are associated with the other two mass values ($m_1$ and $m_5$).  These features can be clearly seen in Fig.~\ref{fig4}, which shows p, the \emph{percentage} difference from the $m_j=0$ curve, versus $E$ for the same three choices of the $U^2_{12}$ parameter.  At $m_1= 4.0 eV$ below the spectrum endpoint Fig.~\ref{fig4} shows a second kink, which was not visible in Fig.~\ref{fig3} because the height of the spectrum was not sufficient so close to the endpoint.  The middle (solid) curve in Fig.~\ref{fig4} also shows error bars which grow in size as the numbers of counts decrease with increasing $E,$ clearly indicating that this second kink in an actual experiment would be much less detectable than the first one.  

Finally, there is a third notable feature in Fig.~\ref{fig4} associated with the $m^2_5=-0.2 keV^2$ or $\mu_5=\sqrt{-m^2_5}=447 eV$ tachyonic mass states.  As can easily be seen from the form of Eq. ~\ref{A1} only $m^2>0$ neutrinos will create kinks, because when $m^2<0$ the expression under the square root is never zero.  The effect of the $m^2<0$ neutrino in Fig.~\ref{fig4} is the sudden rise in p very close to the spectrum endpoint.  The reason for this rise being so close to the endpoint is that the parameter $U^2_{56}$ must be very small given the large negative value of $m^2_5$ and the assumption that $m^2_{eff}=0.$  In fact, of the three spectral features we have discussed, it is very likely that only the first one (the kink at $m_3=21.4 eV$ below the spectrum endpoint) could have shown itself in existing experiments.

\subsection{The Troitsk, Mainz, and Livermore experiments}
Before discussing the third interesting feature of Fig.~\ref{fig4} associated with the $m^2<0$ mass, it is worthwhile considering what the most accurate neutrino mass experiments to date based on tritium $\beta-$decay have observed with respect to the two predicted kinks at 4.0 eV and 21.4 eV.  The experiments by the Troitsk and Mainz groups were conducted in a similar manner during the last two decades, and they have reported similar values for the upper limit on the electron neutrino effective mass.~\citep{Dr2013}  

\subsubsection{Troitsk experiment}  
\begin{figure}[t!]
\centerline{\includegraphics[angle=270,width=1\linewidth]{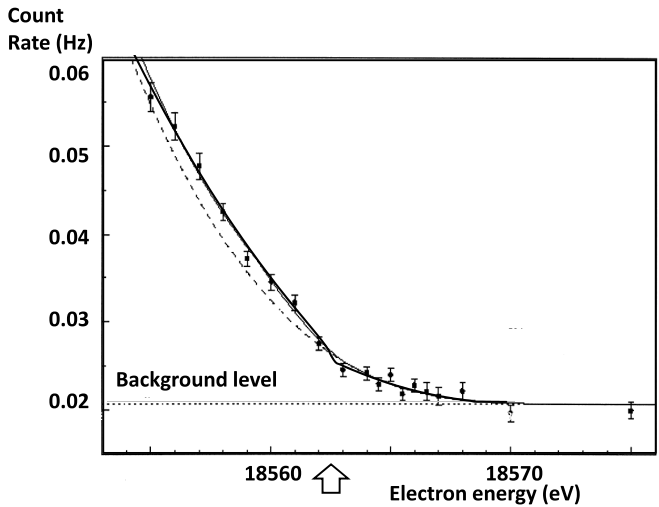}}
\caption{\label{fig5}
{\bf Troitsk data.}  The dashed curve is from ref.~\citep{Lo1999} and it shows the spectrum, after Troitsk authors subtracted their ``step," assuming all $m_j = 0.$ Comparing the data points to the dashed curve yields $\chi^2=44.8$ or $p=4.5\times 10^{-4}.$ The solid curve is the middle curve of Fig.~\ref{fig3} superimposed on the Troitsk data in ref.~\citep{Lo1999} after adjusting the curve's energy scale to give a good fit to the data $\chi^2=12.7$ or $p=0.81$.  It would appear based on this energy scale from ref.~\citep{Lo1999} that the kink occurs about 10 eV before the endpoint (position of upward arrow), but in a later Troitsk article, ref.~\citep{As2012}, that distance was corrected to about 20 eV}
\end{figure}

The one notable difference between the results from the Troitsk and Mainz groups has been a now-disavowed feature known as the ``Troitsk anomaly."   As shown in Fig.~\ref{fig5} in ref.~\citep{Lo1999} a smooth curve through the data would appear to have a kink very similar to one in the solid curves in Fig.~\ref{fig3}.  In order to illustrate how well the middle curve in Fig.~\ref{fig3} fits the Troitsk data we see in Fig.~\ref{fig5} that curve superimposed on the Troitsk data after a suitable adjustment of the scale of the energy axis.  That adjustment was applied because according to Fig.~\ref{fig5}, the anomaly (kink) occurs around 10 eV before the spectrum endpoint, whereas the $3+3$ model prediction is 21.4 eV.  

Despite the good fit ($p=81\%$), how can we justify such an arbitrary adjustment of the energy scale?  In a 2012 publication the Troitsk authors note that based on a reanalysis of their data published 13 years earlier, the location of their anomaly had originally been misplaced.~\citep{As2012}  Fig. 8 in ref.~\citep{As2012} shows their corrected position of the kink for data taken at 9 times of the year.   An average of these 9 data points yields for the corrected kink location the value 19.5 eV before the spectrum endpoint in excellent agreement with the 3 + 3 model prediction.  

The Troitsk authors in their 2012 reanalysis have argued that their anomaly is not really statistically significant after all.~\citep{As2012}  This assertion, however, need not be taken at face value, given the Troitsk author's 13 year long failure to explain this feature initially considered to be highly significant.  A further reason for doubting the Troitsk claim of statistical insignificance is that the 2012 reanalysis included an extra 100 eV spectrum interval at lower energies, which was justified in order to reduce systematic errors.  Such an addition, however, expands greatly the number of bins where a random fluctuation might cause a kink in the spectrum, and this greatly reduces the statistical significance of the anomaly initially seen in 19 energy bins close to the endpoint.  Moreover, the reanalysis did not provide a plot of the data in those original 19 energy bins, so it is unclear if there was any change in the evidence for a kink there, but from the description of what was done in the reanalysis, one would assume the answer to be negative.  Finally, the statistical significance of a spectral kink occurring right at a previously predicted energy (21.4 eV before the endpoint) is far greater than one that occurs at an arbitrary energy.  In conclusion, based on the excellent fit seen in Fig.~\ref{fig5}, and the corrected location of the kink to 19.5 eV based on the reanalysis in ref.~\citep{As2012}, the Troitsk anomaly has the right shape, and occurs at the right location to agree with the $3 + 3$ model prediction.

\begin{figure}[t!]
\centerline{\includegraphics[angle=270,width=1\linewidth]{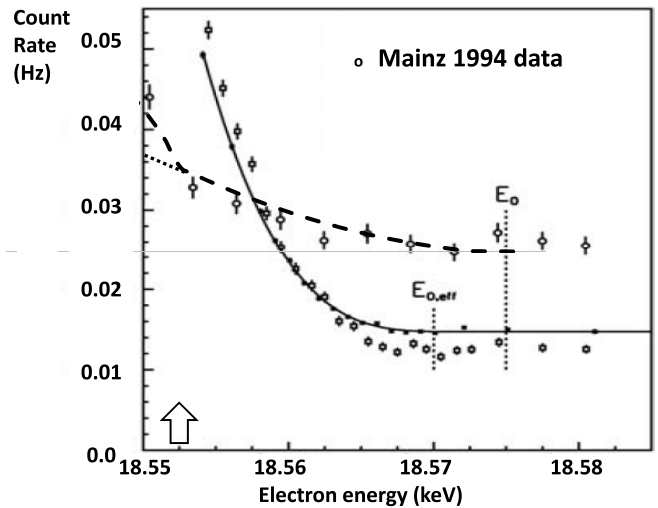}}
\caption{\label{fig7}
{\bf Mainz data.} The three spectra displayed using circles, dots and squares are for data taken by Mainz during the years 1994 (circles), 1998-99 (dots) and 2001 (small squares) from ref.~\citep{Kr2005}.  The solid curve was their m = 0 fit to the 1999-99 data.  The dashed curve has been added to this figure, based on the middle curve from Fig.~\ref{fig3} after adjusting the background level and the vertical scale (but $\emph {not}$ the energy scale) to get a good fit to the 1994 Mainz data.  The up arrow shows the position of the kink.  Note that the horizontal axis in ref.~\citep{Kr2005} was labelled ``retarding potential." }
\end{figure}

\subsubsection{Mainz experiment} The final Mainz results have been reproduced in Fig.~\ref{fig7} taken from Fig. 20 in ref.~\citep{Kr2005}.  The three spectra displayed are for data taken by Mainz during the years 1994 (circles), 1998-99 (black dots) and 2001 (squares).  The 1998-99 data (unlike that for other years) has error bars that are so small as to be invisible on this scale.  We have added to Fig.~\ref{fig7} the middle curve from Fig.~\ref{fig3} (shown dashed here) to fit the 1994 data.  The only free parameters are the background level and the vertical scale.  Clearly the Mainz data for 1994 gives a much better fit to the dashed ``kinky" curve than to the all $m_j=0$ curve (dotted).  A fit to the all $m_j=0$ curve has a $\chi^2 =37.1$ for $p=2.6\times 10^{-5},$ while a fit to the $3+3$ model curve has $\chi^2 = 13.1$ for $p=16\%.$   While the difference in the two fits depends only on a single data point next to the vertical axis, that point lies $>5\sigma$ above the dotted all $m_j=0$ curve.  

The absence of any kink for the Mainz data taken during years other than 1994 has a simple explanation.  As can be seen in Fig.~\ref{fig7}, those data cover only the region 15-20 eV before the endpoint, so they could not possibly provide evidence for any kink at $E_0-E=21.4 eV.$  Thus, in conclusion, despite the Mainz claim of no evidence for a steady state ``Troitsk anomaly" in their data, there is in fact evidence for just such a feature in the Mainz data, and it occurs at an energy consistent with the $3+3$ model prediction.  In fairness to the Mainz authors, however, their assertion of no such anomaly in their data was made before Troitsk reanalyzed their data in 2012 to place the anomaly about 20 eV from the endpoint rather than the original 10 eV.

\subsubsection{Livermore experiment}
\begin{figure}[t!]
\centerline{\includegraphics[angle=270,width=1\linewidth]{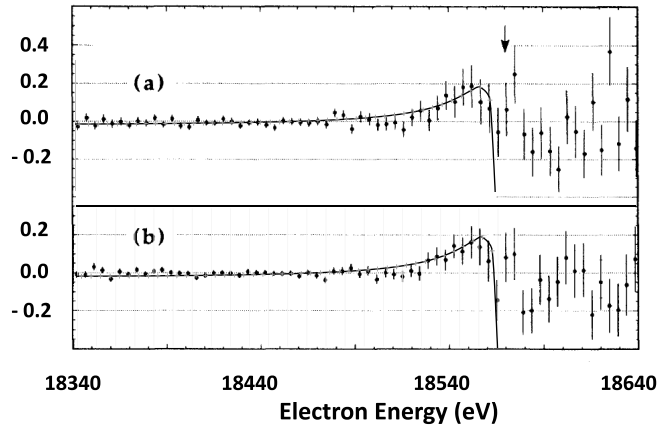}}
\caption{\label{fig8}
{\bf Livermore data.} The data points are ``residuals data" from ref.~\citep{St1995}, i.e., they show the ratio of the data to the all $m_j=0$ fit found in ref.~\citep{St1995} minus 1.0 for two different spectrometer settings corresponding to (a) and (b).  The curves were added by the author.  They were obtained by fitting the difference between the $3+3$ model prediction and the all $m_j=0$ curve having the same normalization.  In doing the fit, the endpoint in the $3 + 3$ model spectrum was shifted by a variable $\Delta E$ relative to the all $m_j=0$ curve until a good fit was obtained (for $\Delta E=3.6 eV$). The arrow shows the fitted spectrum endpoint according to ref.~\citep{St1995}, $E_0=18586.6\pm 25 eV.$}
\end{figure}

A third high statistics experiment done by Stoeffl and Decman at the Lawrence Livermore National Lab was reported in ref.~\citep{St1995}  That 1995 paper reported (a) a limit of 7.2 eV for the $\nu_e$ effective mass, $m_\nu,$ and (b) an anomalous structure in the last 55 eV closest to the endpoint of the tritium spectrum that resulted in a poor fit to the all $m_\nu=0$ case.  In fact, this anomalous structure resulted in a negative value of the effective mass $m_\nu^2= -130 \pm 20 eV^2.$  This $m_\nu^2<0$ result ($6.5\sigma$ below zero) is of course only an artifact, but it shows that the anomalous structure is a highly significant one statistically.  Stoeffl and Decman note that the anomalous structure could be interpreted as a spectral line centered on $23\pm 5 eV$ below the endpoint, folded with the final state distribution, energy resolution function, and energy loss.  The structure shows up most clearly in a plot of the residuals after a fit of their data to the Kurie function $K(E)$ for the all $m_j=0$ case.  This plot (Fig. 2 in ref.~\citep{St1995}) has been reproduced here in Fig.~\ref{fig8}, where the structure appears as a bump or excess of events close to the endpoint shown by the down arrow.  

Superimposed on the Livermore data we show the prediction from the $3 +3$ model, which accounts for the anomalous structure very well. The curves were obtained by calculating the difference between the $3+3$ model prediction (with $U^2_{12}=0.5$) and an all $m_j$ curve, where the endpoint in the former case was shifted by $\Delta E$ relative to the latter.  $\Delta E$ was treated as a free parameter, which was adjusted until a good fit (by eye) to the Stoeffl-Decman residuals plot was obtained.  Thus, the curves shown are not true least squares fits, since it was not possible to extract error bars from the plot in ref.~\citep{St1995}.  The good fit implies that the $3+3$ model describes the actual spectrum, and hence it explains the anomaly centered at $E_0-E=23\pm 5 eV$ -- a value quite consistent with the model prediction of 21.4 eV.  In conclusion, we see that the $3+3$ model kink in the spectrum this distance from the endpoint shows up as a bump in a residuals plot.  Moreover, the finding of a best value for $m_{eff}^2<0$ in this and the other eight high precision  direct mass experiments listed by the Particle Data Group,~\citep{Ol2015} which is a consequence of such a bump near the endpoint, can be attributed to the $3+3$ model, or more specifically the spectral kink occurring 21.4 eV before the endpoint predicted by the model.

\begin{figure}[t!]
\centerline{\includegraphics[angle=270,width=1\linewidth]{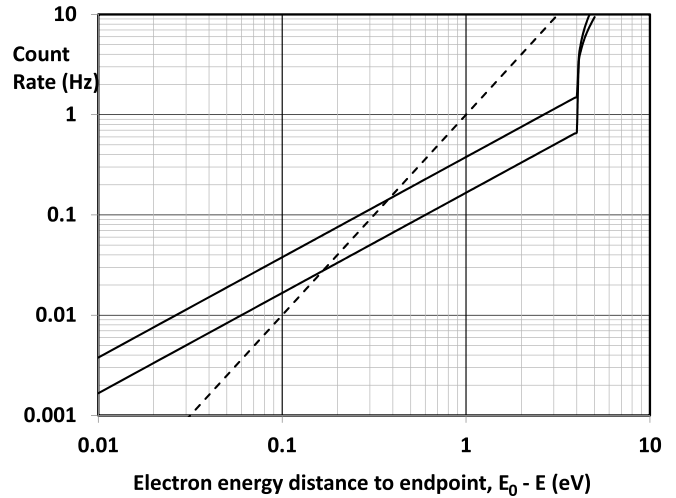}}
\caption{\label{fig6}
The spectra of Fig.~\ref{fig3} shown on a log-log plot for the last 5 eV before the endpoint, here on the left side of the x-axis.  The sloping dashed curve shows the spectrum for the case of all $m_j=0$ and the solid curve corresponds to the $3+3$ model spectrum using $U^2_{12}=0.5.$  The dotted horizontal line is based on ref.~\citep{Ci1999} as discussed in the text.  The arbitrary numbers on the vertical axis are such that one count per second would be observed at $E_0-E=1eV.$}
\end{figure}

\subsection{Signature of the $-0.2 keV^2$ mass state}
The third distinctive spectral feature predicted by the 3 + 3 model is due to the tachyonic mass state.  Fig.~\ref{fig6} (a log-log plot of the middle curve in Fig.~\ref{fig3}) clarifies the nature of this feature.  We see there that the calculated spectrum falls more slowly as the endpoint is approached than would be predicted on the basis of the dashed all $m_j=0$ curve.  Note that endpoint is on the left not right in Fig.~\ref{fig6}.  Based on the figures in ref.~\citep{Dr2013} KATRIN should accumulate about 3 million counts in the last 5 eV of the spectrum during 6 months of data taking.  This should be more than enough data to observe both the predicted kink at 4.0 eV, and also to discriminate between the all $m_j=0$ dashed quadratic curve (for which the count rate N satisfies: $N \propto (E_0-E)^2$) and the solid linear one depicted in Fig. ~\ref{fig6} that vary according to
\begin{equation}
N \propto \mu_5U^2_{56}(E_0-E)
\label{A2}
\end{equation}
where $\mu_5=\sqrt{-m_5^2},$ which follows from the form of Eq.~\ref{A1} when $E_0-E<<\mu_5.$  It should be noted that not only is the form of the spectrum predicted (linear) for the last 4 eV of the spectrum, but so is its normalization.  The normalization can be found using $\mu_5=447 eV,$ and the $U^2_{56}$ value that follows from the $U^2_{12}$ value that gives the best fit to the most prominent predicted kink at $E_0-E=21.4 eV.$

In summary, the KATRIN experiment, if it fulfills its design goals, should provide a definitive test of the $3 + 3$ model by looking for the two predicted kinks occurring at $E_0-E=m_1=4.0 \pm 0.5 eV$ and $E_0-E=m_3=21.4 \pm 1.2 eV$ before the endpoint, and thirdly the linear (rather than quadratic) fall in the spectrum after the kink at 4.0 eV.  If one uses the value for the parameter $U^2_{12}=0.5$ that gave good fits to the Troitsk, Mainz, and Livermore data, the KATRIN data should fit the model spectrum with $\emph{zero}$ adjustable parameters, and the data should exhibit all three predicted features.  

\subsection{Two cautionary notes}
\subsubsection{Functional form of $K^2(E)$ for $m^2<0$ neutrinos}
It has so far been assumed that one can use the same functional form in Eq.~\ref{A1} for both $m^2<0$ and $m^2>0$ neutrinos, which may not be the case.  In some causal tachyon theories, such as that of Ciborowski and Rembielinski,~\citep{Ci1999} or by Radzikowski~\citep{Ra2010}, the functional form of $K^2(E)$ is more complex for $m^2<0$ neutrinos, and it can also be dependent on both the nature of the coupling and the lab velocity relative to a preferred reference frame.  Given as large a tachyonic mass as $\mu_5=447eV$ the shape of the spectra within 4 eV of the endpoint in ref.~\citep{Ci1999} are essentially flat, followed by an abrupt drop to zero at $E=E_0,$ assuming the lab to be close to being the preferred frame -- see the dotted horizontal line in fig.~\ref{fig6}.   Were KATRIN to yield such a result it would be even more distinguishable from the standard all $m_j=0$  curve than the previously discussed linear variation over the last 4.0 eV of the spectrum.

\subsubsection{What if KATRIN refutes the $3+3$ model masses?}
It is possible that the model will be refuted, and the conventional view of the neutrino mass states all having small masses upheld.  In such a case KATRIN still might be able to measure a single effective mass $m_{eff}$ that is significantly different from zero by fitting the spectrum using Eq. ~\ref{A3}.  

\begin{equation}
K^2(E)=(E_0-E)\sqrt{R((E_0-E)^2-m_{eff}^2)}
\label{A3}
\end{equation}

Elsewhere the author has provided evidence for a value for the $\nu_e$ effective mass that is tachyonic:  $m_{eff}^2=-0.11\pm 0.02 eV^2.$~\citep{Eh2015}   If KATRIN fulfills its design goals this ``unphysical" value would be discoverable at the $5\sigma$ level, provided that systematic errors are independent of the sign of the best fit value of $m^2_{eff}.$

\section{Unresolved issues}
This section considers various not-fully-resolved experimental and theoretical issues that might either support or conflict with the $3+3$ model once they are fully resolved.

\subsection{Archived Mainz data on tritium spectrum}  As shown in Fig.~\ref{fig7} the evidence for a kink at 21.4 eV in the Mainz data depended on a single 1994 data point that was $>5\sigma$ above the all $m_j=0$ fitted curve.  If the Mainz data has been archived and is still accessible, it would be most useful if one could inspect the 1994 data for energies further from the endpoint than that one point, and also examine the data for other years in the region 20 to 30 eV before the endpoint, especially the 1998-99 data in view of its very small error bars.

\subsection{Evidence for $3+3$ model from oscillation data?}
The two $m^2>0$ masses of the $3+3$ model might be confirmed by finding evidence for oscillations having: $\Delta m_{13}^2=m_3^2-m_1^2=21.4^2-4.0^2=442 eV^2.$  Such an observation might involve either appearance or disappearance reactions.  Among experiments of the former type we have only upper limits at present with oscillation amplitudes below about $\sin^2 2\theta<10^{-3}$~\citep{As2003} being allowed.   Remarkably, among disappearance experiments there is in fact such a $\Delta m^2 \approx 400eV^2$ value (among many others) currently allowed at the $68\% CL$ having $\sin^2 2\theta>0.1,$ as can be seen in fig. 15 of ref.~\citep{Gi2012}  

\subsection{Consistency of $3+3$ model with oscillation data\label{Osc}?} 
Here it is shown that existing oscillation data are not in obvious conflict with the $3+3$ model.

\subsubsection{Possibility of $\nu \rightarrow \bar{\nu}$ oscillations?}
The possibility of $\nu \rightarrow \bar{\nu}$ oscillations is not an intrinsic part of the $3+3$ model.  It was raised in sect. I as a possible way to justify the small and equal $\Delta m^2/m^2$ for the three $\nu_L,\nu_R$ doublets in the model.  There are, of course, stringent upper limits on the magnitude of active-sterile mixing from both oscillation data.~\citep{Ad2016} and cosmology.~\citep{Sa2015}.   However, if  the members of each $\nu_L, \nu_R$ doublet were a $\nu,\bar{\nu}$ pair this would allow the members of each doublet to both be $\emph {active}$ $\nu$ and $\bar{\nu}$ states, thereby making it much easier to understand the large size of the oscillation amplitudes for $\Delta m^2_{sol}$ and $\Delta m^2_{atm}$ -- much larger than if they involved an active-sterile $\nu_L, \nu_R$ pair.  While it may be true that the KamLAND results effectively rule out the existence of $\nu_e\rightarrow \bar{\nu}_e$ oscillations,~\citep{Ga2012}, that still leaves 15 other possible transitions.  In fact, ref.~\citep{Ho2009} has suggested that the anomalies seen in the LSND and MinnibooNE experiments can be interpreted as being due to $\nu \rightarrow \bar{\nu}$ oscillations, and they note that the MinniBooNE results look more like  $\nu_\mu\rightarrow\bar{\nu}_e$ than $\nu_\mu\rightarrow \nu_e$ events.~\citep{Ho2009} 

\subsubsection{Evidence for there being 3 sterile neutrinos?}
The $3+3$ model consists of 3 $\nu_L, \nu_R$ doublets, which represent either 3 active and 3 sterile neutrinos, or alternatively 3 active neutrinos and 3 active antineutrinos as suggested in the previous section.   Thus, evidence that the number of sterile neutrinos is three would lend support to one variant of the model.  Evidence for sterile neutrinos comes in part from short baseline oscillation experiments, and Conrad et al.~\citep{Co2013} have fit the  data from 14 such experiments for the cases of $3 +N$ neutrinos where $N=0,1,2,3$.  Ref.~\citep{Co2013} reports that the best fit does in fact occur for $N=3,$ and also that some of the fitted sterile neutrino masses are quite large, i.e., $\Delta m^2>>1eV.$ However, those fits were all based on the normal or inverted hierarchy, and specifically it was assumed that $m_1=m_2=m_3=0$ and the mixing between active and sterile mass eigenstates is small.  In contrast, to the 12-parameter fits of ref.~\citep{Co2013}, the $6\times 6$ $U_{i,j}$ matrix applicable to the $3+3$ model involves 25 free parameters (15 mixing angles  $\theta_{i,j}$ and 10 phases $\phi_{i,j}$), which need to fit the transition probabilities between any pair of observed flavor states a and b.   

\subsubsection{Exclusion of an all $\phi_{i,j}=0$ solution in the $3+3$ model}
Finding one or more solutions for the 25 free parameters that agrees with all the observed $P_{a\rightarrow b}$ found from oscillation experiments remains a serious challenge.   It is, however, worth noting that the possibility of a solution existing with all $\phi_{i,j}=0,$ for the $3+3$ model can be ruled out since in this case we have:
\begin{equation}
P_{a\rightarrow b}=\delta_{ab}-A_{ab}sin^2\big(\Delta m^2_{ij}L/4E\big)
\end{equation}
where the amplitudes are given by:
\begin{equation}
A_{ab}=4\Sigma_{i<j} U_{ai}U_{bi}U_{aj}U_{bj}
\end{equation}
which for the special case of oscillations involving $\nu_e$ and $\nu_\mu$ would yield the following relation between the 3 oscillation amplitudes:
\begin{equation}
A_{e\mu}= \sqrt{A_{ee}A{_{\mu\mu}}}
\end{equation}

Specifically, for oscillations corresponding to $\Delta m^2_{atm}=\Delta m^2_{34},$ Eq. 5 predicts: $A_{ee}=4U^2_{e3}U^2_{e4}$, $A_{\mu\mu}=4U^2_{\mu3}U^2_{\mu4},$ 
and $A_{e\mu}=4U_{e3}U_{\mu3}U_{e4}U_{\mu4},$  thus satisfying Eq. 6. However, such a relation between the 3 amplitudes is not consistent with what is actually observed in experiments, and this rules out the all $\phi_{i,j}=0$ possibility for the $3+3$ model.  Of course, Eq. 6 is also not true for the conventional normal $3 + 0$ hierarchy, because there one combines the $\Delta m^2_{13}$ and $\Delta m^2_{23}$ oscillation amplitudes which have nearly the same wavelength.  

In any case, the basic point of this and the previous section is that the large number of free parameters (25) involved in doing a fit when all $\phi_{ij}\ne 0$ is such that one cannot definitively rule out the existence of a solution generally for the $3+3$ model, even though one remains to be found.

\subsection{Consistency of $3+3$ model with cosmology\label{Cosmology}}

\subsubsection{Sum over flavor or mass state masses?}
Based on cosmological models of the early universe there is an equal number density $n_0$ of relic neutrinos of each flavor that depend on the neutrino temperature, $T_\nu$.~\cite{Le2012} These relic neutrinos give rise to a mass/energy density for the $i^{th}$ flavor of $\rho_i=n_0m_i,$ and hence a total energy density of all flavors $\rho_\nu=\Sigma n_0m_i=n_0\Sigma m_i.$  The reason for equal number densities of the flavors is that they are produced in equal numbers from $Z^0$ decay in the early universe.  Thus, the sum of neutrino masses that factors into cosmological models through the neutrino energy density is a sum over $\emph{flavor}$ state (effective) masses, which except for the degenerate case, can be different from the sum over mass state masses.  It is noteworthy that while the sum of mass state and flavor state masses need not be equal, unitarity requires that the sum of their squares are identical:
\begin{equation}
\Sigma_i m^2_i=\Sigma_i \Sigma_j |U_{ij}|^2m_j^2=\Sigma_j m_j^2\approx -0.2 keV^2
\label{Eq7}
\end{equation}
This identity thus requires that at least one flavor state have a non-negligible tachyonic mass, possibly as large as $m^2\approx -0.2 keV^2.$  Presumably, such a flavor state would need to be a sterile one, with negligible mixing with the active states to achieve consistency of observations of neutrino oscillations and cosmology.

\subsubsection{Definition of the effective mass in cosmology}
It is plausible that flavor states produced in $Z^0$ decay have an effective mass defined as in (single) $\beta-$decay, although for $\beta-$decay a single effective mass only characterizes the shape of the $\beta-$spectrum in an experiment if the individual masses $m_j$ are sufficiently degenerate.  The relation: $m^2_{i(eff)}=\Sigma |U_{ij}|^2m_j^2$ follows from the definition of the expectation value of the operator $m^2=E^2-p^2$ calculated for the $i^{th}$ flavor state
$|\nu_i>=|\Sigma U_{ij}\nu_j>$, namely 
\begin{equation}
m^2_{i(eff)}=<\Sigma_jU_{ij}\nu_i |m^2|\Sigma_kU_{ik}\nu_k>=\Sigma |U_{ij}|^2m_j^2
\end{equation}

Note that the validity of the $m_{(eff)}$ equation does not depend on the near-degeneracy of the $m_j.$  As a consequence, if two mass eigenstate doublets are very heavy and have $m^2>0$ as in the $3+3$ model, the third doublet must have $m^2<0.$  Only in this way could all flavor states have $m^2_{i(eff)}<< 1 eV$ (assuming a suitably chosen $U_{ij}$), and thereby allow these states to satisfy the cosmological constraint on their sum: $m_{tot}=\Sigma m_i <0.3 eV$.~\citep{Ol2015} Based on the preceding discussion, the $3+3$ model plausibly could be consistent with the cosmological limit on the sum of the neutrino flavor state (effective) masses.

\subsubsection{Tachyonic flavor states in the mass sum, $\Sigma m$}
Let us finally consider how tachyonic flavors might impact cosmology.   It has long been known that tachyons can have a negative energy,~\citep{Dh1968}, and that negative energy density offers a simple way to explain dark energy,~\citep{No2005}, one form of which might involve a sea of tachyonic neutrinos.~\citep{Je2013}.  A negative energy density $\rho_i=n_0m_i$ for tachyonic neutrinos of the $i^{th}$ flavor requires that their gravitational neutrino mass $m_i$ be considered to be negative (not imaginary) since their spatial number density $n_0$ cannot be.  Thus, as suggested in ref.~\citep{Eh2015}, for a non-sterile tachyonic flavor mass $m^2<0$ we use $\mu=-\sqrt{|m^2|}$ in the sum over flavor state masses.\\  

\section{Summary}

The 3 + 3 model of the neutrino mass eigenstates proposed by the author in 2013 postulated three $\nu_L, \nu_R$ doublets having specific unconventionally heavy masses, one of which has $m^2 < 0$. It is shown that the model is not in obvious violation of empirical constraints from cosmology and particle physics, including direct mass experiments based on the shape of the tritium $\beta-$decay spectrum near its endpoint.   In fact, despite the small (2eV) upper limit these experiments set on the $\nu_e$ effective mass, it is shown that excellent fits can be obtained to the much larger model masses. 

To calculate the $\beta-$spectrum predicted by the three large nondegenerate masses of the model it is necessary to use Eq.~\ref{A1}, and not Eq.~\ref{A3} based on a single effective mass.  Nevertheless, an auxiliary condition is imposed that $m_{eff}\approx 0,$ so as to comply with cosmological constraints.    As a result, the predicted shape of the spectrum depends on only one free parameter, $U^2_{12}\equiv U^2_{e1}+U^2_{e2},$ which is the weight of the first of the three masses (4.0 eV) in Eq.~\ref{A1}.   The calculated spectrum has three ``anomalies" compared to that for all $m_j=0$, and the only one prominent enough to be seen with existing data would be a kink in the spectrum at a distance from the endpoint, $E_0-E=m=21.4\pm 1.2 eV,$ the second of the three model masses.  Data from the three most precise tritium experiments to date by the Troitsk, Mainz and Livermore collaborations, each show evidence for this feature $\emph{and with a common weight ($U^2_{12}=0.5$)}.$

The good fits of the $3+3$ model to these three data sets (much better than the fits to the all $m_j=0$ case), make for a total of six pieces of evidence for its unconventionally large $m^2>0$ masses.  The earlier evidence includes the SN 1987A data fit that gave rise to the model~\citep{Eh2012,Eh2013}, the dark matter fit to (a) the Milky Way (see fig. 3 in ref.~\citep{Ch2014}), and (b) that for four clusters of galaxies (see fig. 1 in ref.~\citep{Ch2014}). Moreover, one can also find validation for the existence of a $m^2<0$ neutrino mass state, in light of evidence presented in ref.~\citep{Eh2015} for a tachyonic flavor state ($\nu_e$), which is required by Eq.~\ref{Eq7}.  Given this volume of support for the model, there are reasons to suspect the KATRIN experiment just might validate it, including the $m^2<0$ mass.  Even if the theoretical basis of the model is speculative, skeptical readers should not forget that much of what is known about neutrinos was initially suggested on a speculative basis, including their very existence~\citep{Br1976}.\\

\section*{Acknowledgments} The author is immensely grateful to Alan Chodos for his encouragement and for his numerous helpful comments on various iterations of the manuscript and to all the anonymous reviewers of this and earlier drafts whose severe criticisms may have delayed the manuscript's publication, but also resulted in a greatly strengthened final version.

\end{document}